\documentclass[11pt]{cernrep}
\usepackage{graphicx,epsfig}
\newcommand{\be}{\begin{equation}}
\newcommand{\ee}{\end{equation}}
\newcommand{\bea}{\begin{eqnarray}}
\newcommand{\eea}{\end{eqnarray}}

\newcommand{\as}{$\alpha_s$}
\def\lsim{\mathrel{\mathpalette\@versim<}}
\def\gsim{\mathrel{\mathpalette\@versim>}}
 \def\@versim#1#2{\lower0.2ex\vbox{\baselineskip\z@skip\lineskip\z@skip
       \lineskiplimit\z@\ialign{$\m@th#1\hfil##$\crcr#2\crcr\sim\crcr}}}
\catcode`@=12 

\newcommand{\ben}{\begin{enumerate}}
\newcommand{\een}{\end{enumerate}}

\def\frac#1#2{{{#1}\over {#2}}}
\def\gsim{\mathrel{\rlap{\lower4pt\hbox{\hskip1pt$\sim$}}
    \raise1pt\hbox{$>$}}}         
\def\lsim{\mathrel{\rlap{\lower4pt\hbox{\hskip1pt$\sim$}}
    \raise1pt\hbox{$<$}}}         

\newcommand{\draft}[1]{}

\def \n0{N_j^{(0)}}

\def\lapprox{\lower .7ex\hbox{$\;\stackrel{\textstyle <}{\sim}\;$}}
\def\gapprox{\lower .7ex\hbox{$\;\stackrel{\textstyle >}{\sim}\;$}}

\begin{document}

\title{\boldmath The PDF4LHC Working Group Interim Recommendations}

\author{
Michiel Botje$^1$,
Jon Butterworth$^2$,
Amanda Cooper-Sarkar$^3$,
Albert de Roeck$^4$,
Joel Feltesse$^{5}$,
Stefano Forte$^{6}$,
Alexander Glazov$^{7}$,
Joey Huston$^{8}$,
Ronan McNulty$^{9}$,
Torbj\"orn Sj\"ostrand$^{10}$,
Robert ~S. ~Thorne$^2$
}


\institute{
$^1$ NIKHEF, Science Park, Amsterdam, The Netherlands\\
$^2$ Department of Physics and Astronomy, University College, London, WC1E 6BT, UK\\
$^3$ Department of Physics, Oxford University, Denys Wilkinson Bldg, Keble Rd, Oxford, OX1 3RH, UK\\
$^4$ CERN, CH--1211 Gen\`eve 23, Switzerland; Antwerp University, B--2610 Wilrijk, Belgium; University of California Davis, CA, USA\\
$^{5}$ CEA, DSM/IRFU, CE-Saclay, Gif-sur-Yvetee, France\\
$^{6}$ Dipartimento di Fisica, Universit\`a di Milano and INFN, Sezione di Milano, Via Celoria 16, I-20133 Milano, Italy\\
$^{7}$ Deutsches Elektronensynchrotron DESY Notkestra{\ss}e 85 D--22607 Hamburg, Germany \\
$^{8}$ Physics and Astronomy Department, Michigan State University, East Lansing, MI 48824, USA\\
$^{9}$ School of Physics, University College Dublin Science Centre North, UCD Belfeld, Dublin 4, Ireland\\
$^{10}$ Department of Astronomy and Theoretical Physics, Lund University, S\"olvegatan 14A, S-223 62 Lund, Sweden\\
}

\maketitle

\begin{abstract}

This note
provides an interim summary of the current
recommendations of the PDF4LHC working group for the use of parton
distribution functions (PDFs) and of PDF uncertainties at the LHC, for cross
section and cross section uncertainty calculations. 
It also contains a succinct user guide to the
computation of PDFs, uncertainties and correlations using available
PDF sets.

A companion note (the PDF4LHC Working Group Interim Report) summarizes predictions for benchmark cross sections at the LHC at NLO using modern PDFs currently available from 6 PDF fitting groups. 

\end{abstract}

\clearpage

\tableofcontents

\clearpage

\section{Introduction}
\label{intro}

The LHC experiments are currently producing cross sections from the 7 TeV data, and thus need accurate predictions for these cross sections and their uncertainties at NLO and NNLO. Crucial to the predictions and their uncertainties are the parton distribution functions (PDFs) obtained 
from global fits to data from deep-inelastic scattering, Drell-Yan and
jet data. A number of groups have produced publicly available PDFs
using different data sets and analysis frameworks. It is one of the
charges of the PDF4LHC working group to evaluate and understand
differences among the PDF sets to be used at the LHC, and to provide a
protocol for both experimentalists and theorists to use the PDF sets
to calculate central cross sections at the LHC, as well as to estimate
their PDF uncertainty. This note is intended to provide recommendations for estimations of cross sections and uncertainties at the LHC. 

\section{The PDF4LHC recommendation}
\label{sec:pdf4lhcreco}

Before the recommendation is presented, it is useful to highlight the differences between two use cases: (1) cross sections which have not yet been measured (such as, for example, Higgs production) and (2) comparisons to existing cross sections. For the latter, the most useful comparisons should be to the predictions using individual PDFs (and their uncertainty bands). Such cross sections have the potential, for example, to provide information useful for modification of those PDFs. For the former, in particular the cross section predictions in this report, we would like to provide a reliable estimate of the true uncertainty, taking into account possible differences between the central values of predictions using different PDFs~\footnote{It may also be more conservative to use this procedure to calculate the uncertainty for the acceptance for a measured cross section.}. From the results seen it is clear that this uncertainty will be larger than that from any single PDF set, but we feel it should not lose all connection to the individual PDF uncertainties (which would happen for many processes if the full spread of all PDFs were used), so some compromise is proposed.

The wish for a recommendation follows directly from the HERALHC workshop conclusions~\cite{heralhc}, and has always been one of the main goals of the PDF4LHC group since its creation in 2006, particularly as a wish of the LHC experiments. In order for the recommendation to be acceptable by the experiments, it  has to be pragmatic and not unnecessarily complicated. It is also an advantage to try to keep close to the techniques or procedures that are already being used in the experiments up to now. To that end, many studies in the past years have been done with CTEQ and MSTW, and recently also NNPDF.  But it should be clear that at this point no general judgement is made on whether certain PDFs can or cannot be used; for any given particular  analysis, different expert judgements can lead to  different choices, maybe even the use of only a single PDF set. Also, the recommendation given below can and will be revised in due time when a new level of understanding and development is reached, which is expected to  follow from the ongoing discussions at the PDF4LHC forum.
  
As seen at NLO there is always reasonable agreement between MSTW, CTEQ and NNPDF and potentially more deviation with the other sets. In some cases this deviation has at least one potential origin, e.g. the $\bar t t$ cross-section at 7~TeV at the LHC probes similar PDFs as probed in the lower-$p_T$ jet production at the Tevatron, which has neither been fit nor validated against quantitatively by some groups (preliminary results for ABM may be found at \cite{Trento}). As noted, large deviations in predictions between existing NNLO sets are similar to those between the same NLO sets. Discrepancies in MSTW, CTEQ and NNPDF do not always have clear origin, or may be a matter of procedure (e.g. gluon parameterisation) which is an ongoing debate between groups. Bearing this in mind and having been requested to provide a procedure to give a moderately conservative uncertainty, PDF4LHC recommends the following.

\subsection{NLO prescription}

At NLO, the recommendation is, for the first case  described above, to use predictions from the PDF fits from CTEQ, MSTW and NNPDF. These sets all use results from hadron collider experiments, i.e. the Tevatron, as well as fixed target experiments and HERA, and they make available specific sets for a variety of 
$\alpha_s(m_Z)$ values. 
The PDFs from these three groups to be used are: CTEQ6.6~\cite{Nadolsky:2008zw},
MSTW2008~\cite{Martin:2009iq} and NNPDF2.0~\cite{Ball:2010de}.  Neither the CTEQ6.6 nor the MSTW2008  PDF sets
use the new combined very accurate HERA data sets, which are instead
used by NNPDF2.0 (updates of the CTEQ (CT10~\cite{Lai:2010vv}) and
MSTW~\footnote{The MSTW presentation at the DIS 2010 workshop~\cite{Thorne:2010kj} can be
  consulted to assess the effects of these data.}   PDFs will include
them) but these PDFs are  now most commonly used by the LHC
experiments and are suggested in the recommendation for this
reason. The NNPDF2.0 set does not use a general-mass variable flavor
number scheme (the NNPDF2.1 PDF set, which does use a general-mass
variable flavor number scheme is currently being
finalized~\footnote{The NNPDF presentation at the DIS 2010
  workshop~\cite{Rojo:2010gv} can be
  consulted to assess the effects of these corrections.}), but the
alternative method which NNPDF use for determining PDF uncertainties
provide important independent information.  

Other PDF sets, ABKM09~\cite{Alekhin:2009ni,Alekhin:2010iu} GJR08~\cite{Gluck:2007ck,Gluck:2008gs} and HERAPDF1.0~\cite{herapdf10} are useful for direct comparison to data as suggested for case (2), for cross
checks, and for a more extensive and conservative evaluation of the PDF uncertainty. For example, HERAPDF1.0 allows a study of the theoretical uncertainties related to the charm mass treatment.
The  $\alpha_s$ uncertainties (for the PDFs) can be evaluated by
taking a range of $\pm 0.0012$ for $68\%$c.l. (or $\pm 0.002$ for $90\%$
c.l.) from the preferred central value for CTEQ and NNPDF. The total
PDF+$\alpha_s$ uncertainty can then be evaluated by adding the
variations in PDFs due to $\alpha_S$ uncertainty in quadrature with
the fixed $\alpha_S$ PDF uncertainty (shown to correctly incorporate
correlations in the quadratic error approximation~\cite{Lai:2010nw}) or, for NNPDF,
more efficiently taking a gaussian distribution of PDF replicas
corresponding to different values of $\alpha_s$~\cite{Demartin:2010er,LH}. 
For MSTW the PDF+$\alpha_s$ uncertainties should be evaluated using
their prescription which better accounts for correlations between the
PDF and $\alpha_s$ uncertainties when using the MSTW dynamical
tolerance procedure for uncertainties~\cite{Martin:2009bu}. Adding the $\alpha_S$
uncertainty in quadrature for MSTW can be used as a simplification but
generally gives slightly smaller uncertainties.  

So the prescription for NLO is as follows:

\begin{itemize}

\item For the  calculation of uncertainties at the LHC, use  the
  envelope provided by the central values and PDF+$\alpha_s$ errors
  from the MSTW08, CTEQ6.6 and NNPDF2.0 PDFs, using each group's
  prescriptions for combining the two types of errors. We propose this
  definition of an envelope because the deviations between the
  predictions are as large as  their uncertainties. 
  As a central value, use the midpoint of this
  envelope. We recommend that a 68\%c.l. uncertainty envelope be
  calculated and the $\alpha_s$ variation suggested is consistent with
  this. Note that the CTEQ6.6 set has uncertainties and $\alpha_s$
  variations provided only at 90\%c.l. and thus their uncertainties
  should be reduced by a factor of 1.645 for 68\%c.l.. Within the
  quadratic approximation, this procedure is completely correct.

\end{itemize}

\subsection{NNLO prescription}

At NNLO, base the calculation of PDF uncertainties on the only NNLO set which currently includes a wide variety of hadron collider data sets, i.e. MSTW2008~\footnote{Although inclusive jet data from the Tevatron are included in the MSTW2008 (and other) NNLO fits, we note that to date the inclusive jet cross section has only been cslculated to NLO.}. 
There seems to be
no reason to believe that the spread in predictions of the global fits, i.e. MSTW, CTEQ and NNPDF,  will diminish significantly at NNLO compared to NLO, where this spread was somewhat bigger than the uncertainty from each single group. 
Hence, at NNLO the uncertainty obtained from MSTW alone should be expanded 
to some degree. It seems most appropriate to do this by multiplying the MSTW 
uncertainty at NNLO by the factor obtained by dividing the full uncertainty 
obtained from the envelope of MSTW, CTEQ and NNPDF results at NLO by the MSTW 
uncertainty at NLO. In all cases the $\alpha_s$ uncertainty should be included. We note that in most cases so far examined for the LHC running at 7TeV centre of mass energy this factor of the envelope divided by the MSTW uncertainty is quite close to 2, and this factor can be used as a short-hand prescription. 

Since there are NNLO PDFs obtained from fits by the ABM,GJR~\footnote{See Ref.~\cite{Alekhin:2010dd} for a comparison of a number of benchmark cross sections at NNLO.} and HERAPDF groups, these should ideally be compared with the above procedure. 

So the prescription at NNLO is:

\begin{itemize}

\item As a central value, use the MSTW08 prediction. As an uncertainty, take the same percentage uncertainty on this NNLO prediction as found using the  NLO uncertainty prescription given above.

\end{itemize}

\clearpage

\section{The PDF4LHC prescription for Higgs production via gluon fusion}
\label{sec:pdf4lchiggs}

The total hadronic cross-section
for the inclusive production of a Higgs boson via gluon fusion
can be written as
\be
\sigma(h_1 h_2\to H+X)
=
\sum_{a,b}\int_0^1 dx_1 dx_2~
f_{a,h_1}(x_1,\mu_F^2) f_{b,h_2}(x_2,\mu_F^2)
~
\int_0^1 dz~
\delta\left(
z-\frac{\tau_H}{x_1 x_2}
\right)
\hat\sigma_{ab}(z)
\ee

The accurate evaluation of
the partonic cross section $\hat \sigma_{ab}$
is the most important step to obtain an accurate prediction of
the central value of the total hadronic cross section.
In fact the latter receives large contributions from NLO- and NNLO-QCD
corrections, but also from the soft gluon resummation, from leading
NNNLO-QCD terms and from NLO-EW corrections.
Finite top mass corrections at NNLO-QCD have been studied and turn out
to be small.
For the sake of simplicity but still in full generality,
the evaluation of the size of the combined PDF+\as uncertainty
can be obtained by considering only exact NLO-QCD and NNLO-QCD corrections
in the infinite top mass limit.
The results in the next Sections have been computed using the code described in
\cite{ABDV}, improved with the NNLO-QCD corrections~\cite{nnlopapers}.

In this study~\footnote{We would like to thank Alessandro Vicini for carrying out this study.} the following three global sets of PDF have been
considered:
CTEQ6.6~\cite{Nadolsky:2008zw}, 
MSTW2008~\cite{Martin:2009iq, Martin:2009bu}, 
NNPDF2.0~\cite{Ball:2010de}. The combined PDF+$\alpha_s$ uncertainties
for each of the three global sets is computed as discussed in
the PDF4LHC Working Group Interim Report.
The various recipes
are also compared in detail in Ref.~\cite{Demartin:2010er}.

In order to obtain a meaningful comparison of the predictions computed
with different PDF sets, it is crucial to adopt the same
uncertainty range for the \as value. For the purposes of illustration
of the PDF4LHC recommendation, we will assume the same range as
the one used for the PDF4LHC Working Group Interim Report
namely,
\be
\delta^{(90)}\alpha_s=0.002~~~90~\%c.l.,\quad\quad
\delta^{(68)}\alpha_s=0.0012~=~0.002/C_{90}~~~68~\%c.l.
\ee
where $C_{90}=1.64485$ is the number of standard deviations
corresponding to a $90\% ~c.l.$

At NLO-QCD, the PDF4LHC recipe can be summarized in the following steps.
\begin{enumerate}
\item
Compute the Higgs cross section, using CTEQ6.6, MSTW2008nlo68cl, NNPDF2.0.
\item
For each set, use the preferred $\alpha_s(m_Z)$ value (respectively 0.118,  0.1207, 0.119).
\item
Compute the PDF+\as uncertainty band, according to the rules of
each collaboration described in the PDF4LHC Working Group Interim Report 
\item
Take the envelope of the three uncertainty bands.
\item
Compute the mid-point of the resulting band
and call uncertainty the distance of the edge of the envelope from it.
\end{enumerate}

\begin{figure}[ht]
\begin{center}
\includegraphics[height=65mm,angle=0]{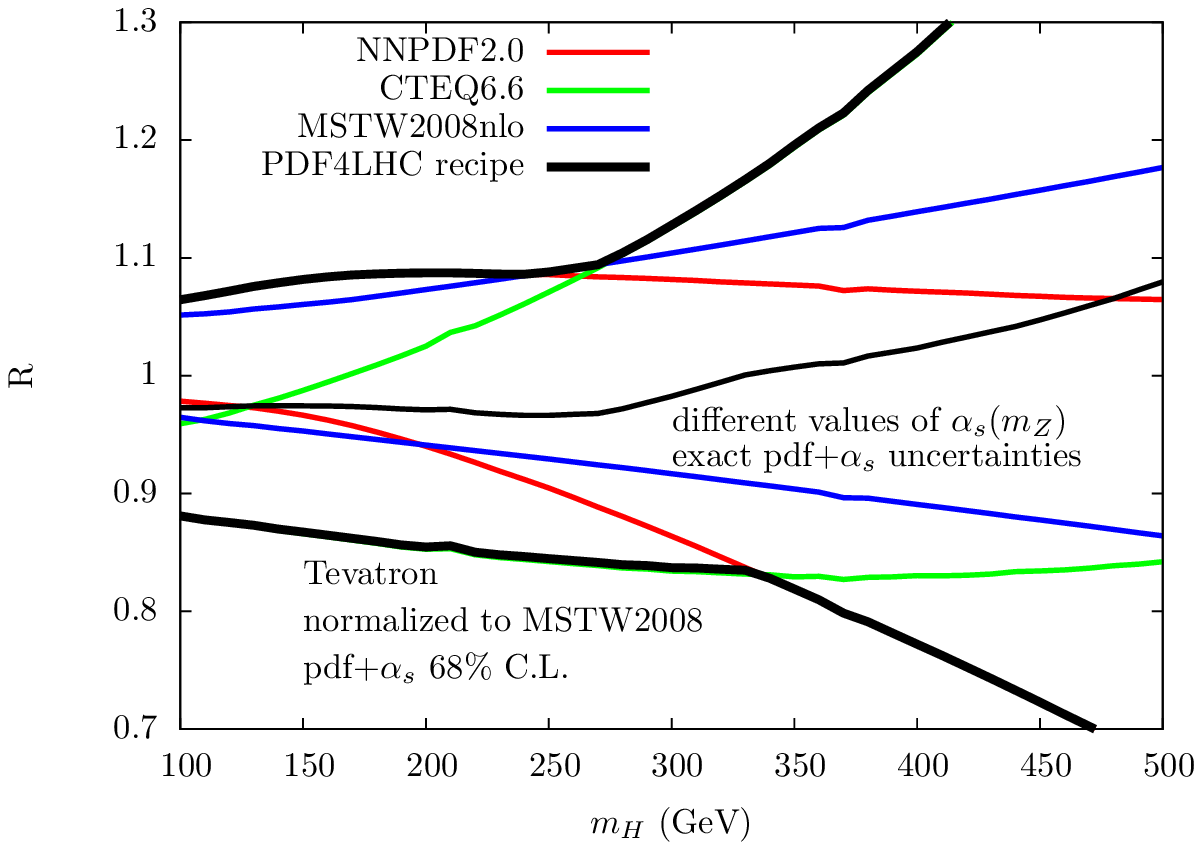}\\
\includegraphics[height=65mm,angle=0]{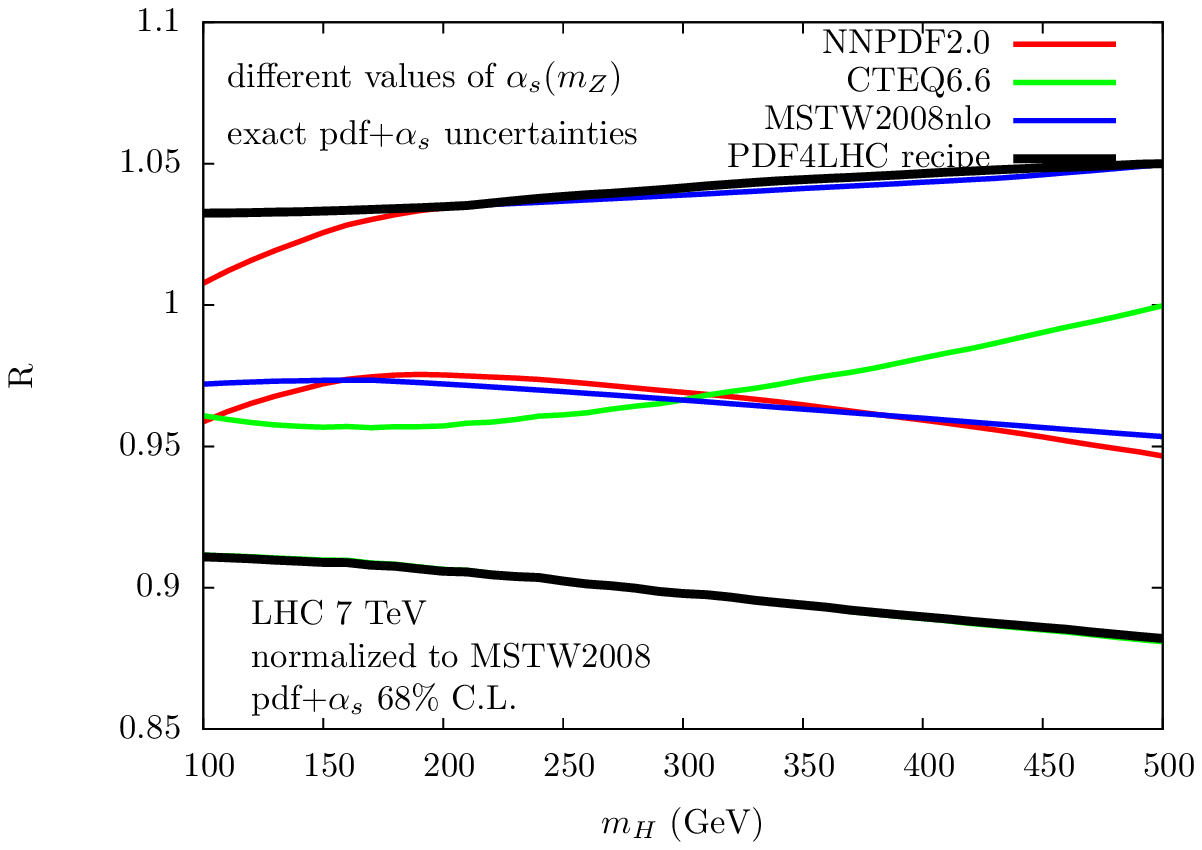}\\
\includegraphics[height=65mm,angle=0]{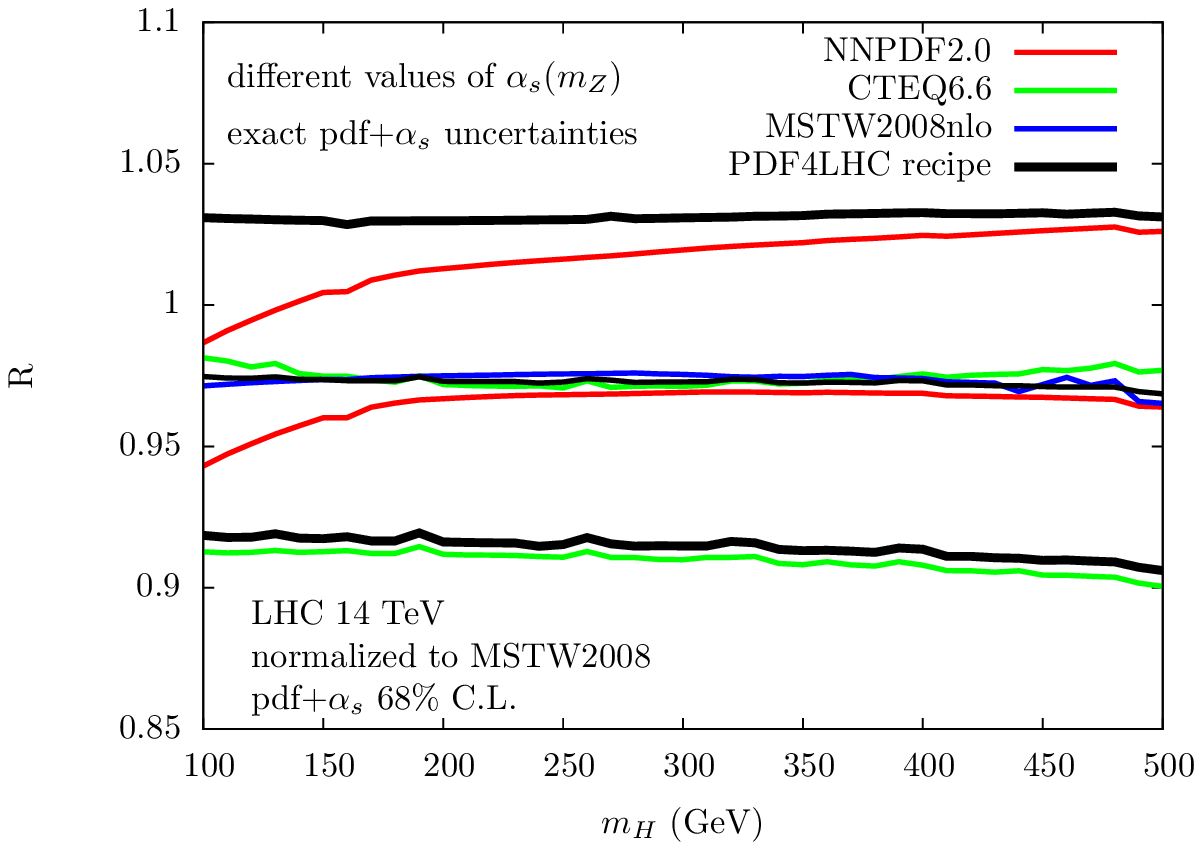}
\end{center}
\caption{Combined PDF+\as uncertainty band
relative to the total Higgs production cross section via gluon fusion,
at NLO-QCD,
evaluated according to the PDF4LHC recipe.
The bands are normalized with the central value of MSTW2008nlo.}
\label{nloenvelope}
\end{figure}

In Fig.~\ref{nloenvelope} we show at different collider energies
(Tevatron, LHC 7 and 14 TeV)
the size of the combined PDF+\as uncertainty bands obtained with
CTEQ6.6, MSTW2008nlo68cl, NNPDF2.0, all normalized to the central
value by MSTW2008nlo68cl.
One remarks that for different Higgs mass values the predictions show
partial agreement of different pairs of the three collaborations, 
in such a way that only an envelope (the black line)
of the three bands provides 
a conservative estimate of the uncertainty in the whole mass spectrum.

\begin{figure}[ht]
\begin{center}
\includegraphics[height=65mm,angle=0]{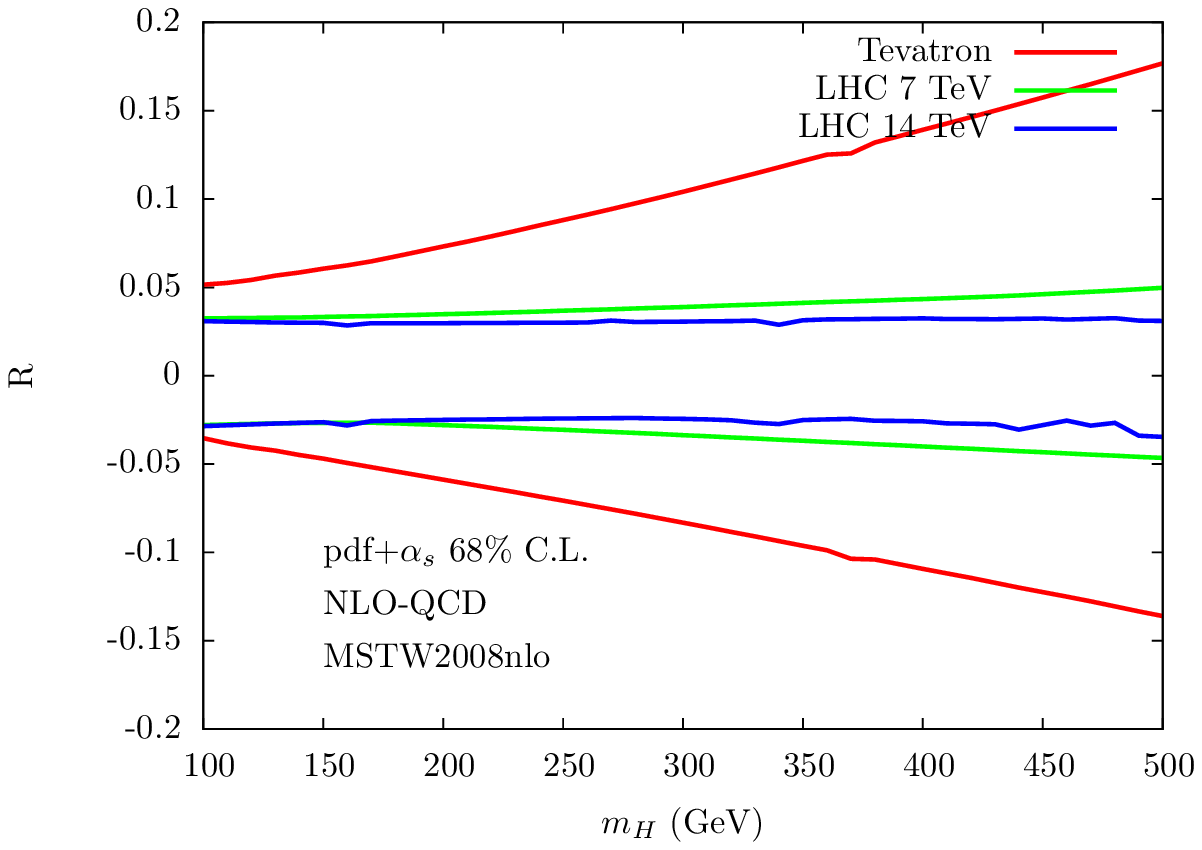}
\includegraphics[height=65mm,angle=0]{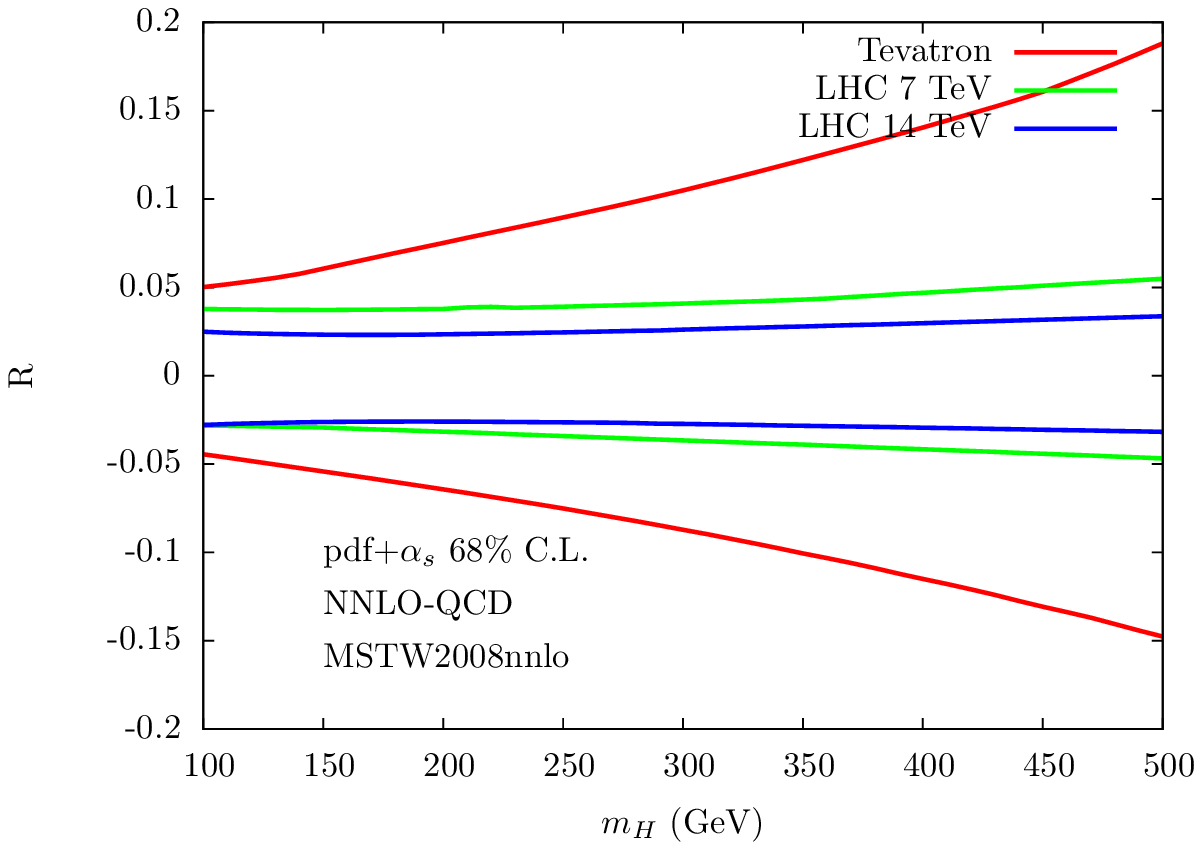}\\[5mm]
\includegraphics[height=65mm,angle=0]{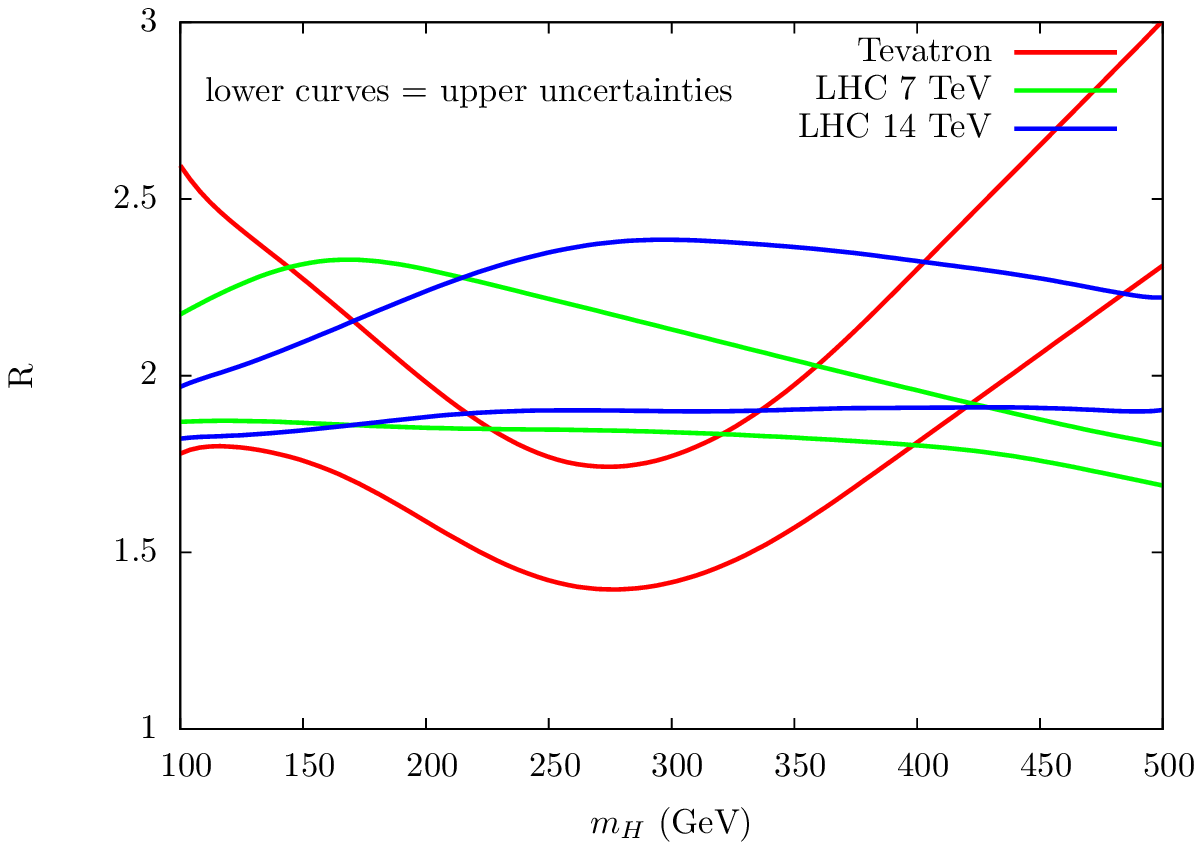}
\end{center}
\caption{In the upper panels, combined PDF+\as uncertainty band
relative to the total Higgs production cross section via gluon fusion,
at NLO-QCD and at NNLO-QCD, obtained with MSTW2008nlo68cl and with
MSTW2008nnlo68cl.
In the lower panel,
rescaling factor obtained from the ratio of the percentage
  width of the NLO-QCD envelope with respect to its mid point
  to the percentage uncertainty of the MSTW2008nlo68cl band.
The rescaling factor has to be applied to the MSTW2008nnlo PDF+\as
band, to obtain an estimate of the NNLO-QCD envelope.}
\label{nlonnlo}
\end{figure}

Now we turn to discuss NNLO-QCD.
There is, at present, only one global set of PDF
extracted with NNLO-QCD accuracy: MSTW2008nnlo.
In this case it is not possible to prepare an envelope like it is done
at NLO-QCD.
The PDF4LHC recipe in this case extrapolates the information available
at NLO-QCD to NNLO and
can be summarized in the following steps.
\begin{enumerate}
\item
From the NLO-QCD exercise determine the percentage uncertainty 
       relative to the mid point of the envelope.
\item
Determine the ratio of this percentage uncertainty to the one
estimated with MSTW2008nlo68cl.
\item
Compute the central value and the PDF+\as uncertainty band
with MSTW2008nnlo68cl.
\item
Rescale the MSTW2008nnlo percentage uncertainty by the above ratio.
\end{enumerate}
One remarks from Fig.~\ref{nlonnlo} 
that the PDF+\as uncertainty bands 
obtained with MSTW2008 at NLO- and at NNLO--QCD
are very similar. 
The small differences are taken into account by rescaling the
MSTW2008nnlo68cl uncertainty band
with the ratio at NLO-QCD of the percentage width of the envelope with
respect to its mid point, to the 
percentage uncertainty of MSTW2008nlo68cl.
As it can be observed in Fig.~\ref{nlonnlo} (lower panel),
the rescaling factor is close to 2, but it has a non trivial
dependence on the Higgs boson mass, on the collider type and on the
collider energy.

\clearpage

\section{Summary}
\label{sec:summary}

In this note,
we have provided a method to calculate reasonable estimates
for PDF uncertainties for cross section predictions at the LHC, using PDFs from CTEQ, MSTW and NNPDF. The method is intended to be both pragmatic and  practical for the calculation of PDF uncertainties; it is not intended to discourage comparisons to all PDFs relevant for LHC calculations. 

The recommendation is expected to evolve when new experimental sets
and new PDF determinations warrant. 

\clearpage



\begin{thebibliography}{99}

\bibitem{heralhc} http://www.desy.de/\~heralhc/

\bibitem{Trento} http://indico.cern.ch/conferenceDisplay.py?confId=93790

\bibitem{Nadolsky:2008zw}
  P.~M.~Nadolsky {\it et al.},
  Phys.\ Rev.\  D {\bf 78} (2008) 013004
  [arXiv:0802.0007 [hep-ph]].

\bibitem{Martin:2009iq}
  A.~D.~Martin, W.~J.~Stirling, R.~S.~Thorne and G.~Watt,
  Eur.\ Phys.\ J.\  C {\bf 63} (2009) 189
  [arXiv:0901.0002 [hep-ph]].

\bibitem{Ball:2010de}
  R.~D.~Ball, L.~Del Debbio, S.~Forte, A.~Guffanti, J.~I.~Latorre, J.~Rojo and M.~Ubiali,
  Nucl.\ Phys.\  B {\bf 838}, 136 (2010)
  [arXiv:1002.4407 [hep-ph]].

\bibitem{Lai:2010vv}
  H.~L.~Lai, M.~Guzzi, J.~Huston, Z.~Li, P.~M.~Nadolsky, J.~Pumplin and C.~P.~Yuan,
Phys.Rev.D {\bf 82}, 074024 (2010).
  arXiv:1007.2241 [hep-ph].


\bibitem{Thorne:2010kj}
  R.~S.~Thorne, A.~D.~Martin, W.~J.~Stirling and G.~Watt,
  arXiv:1006.2753 [hep-ph].

\bibitem{Rojo:2010gv}
  J.~Rojo {\it et al.},
  arXiv:1007.0354 [hep-ph].

\bibitem{Alekhin:2009ni}
  S.~Alekhin, J.~Bl\"umlein, S.~Klein and S.~Moch,
  Phys.\ Rev.\  D {\bf 81} (2010) 014032
  [arXiv:0908.2766 [hep-ph]].

\bibitem{Alekhin:2010iu}
  S.~Alekhin, J.~Blumlein and S.~Moch,
PoS D {\bf IS2010}, 021 (2010)
  arXiv:1007.3657 [hep-ph].


\bibitem{Gluck:2007ck}
  M.~Gluck, P.~Jimenez-Delgado and E.~Reya,
  Eur.\ Phys.\ J.\  C {\bf 53} (2008) 355
  [arXiv:0709.0614 [hep-ph]].

\bibitem{Gluck:2008gs}
M.~Gluck, P.~Jimenez-Delgado, E.~Reya and C.~Schuck,
Phys. Lett. B {\bf 664}, 133 (2008)
[arXiv:0801.3618 [hep-ph]].

\bibitem{herapdf10}
  F.~D.~Aaron {\it et al.}  [H1 Collaboration and ZEUS Collaboration],
  JHEP {\bf 1001} (2010) 109
  [arXiv:0911.0884 [hep-ex]].



\bibitem{Lai:2010nw}
  H.~L.~Lai, J.~Huston, Z.~Li, P.~Nadolsky, J.~Pumplin, D.~Stump and C.~P.~Yuan
  Phys.\ Rev.\ D {\bf 82}, 054021 (2010)
  arXiv:1004.4624 [hep-ph].

%





\bibitem{Demartin:2010er}
  F.~Demartin, S.~Forte, E.~Mariani, J.~Rojo and A.~Vicini,
  Phys.\ Rev.\  D {\bf 82} (2010) 014002
  [arXiv:1004.0962 [hep-ph]].





\bibitem{Martin:2009bu}
  A.~D.~Martin, W.~J.~Stirling, R.~S.~Thorne and G.~Watt,
  Eur.\ Phys.\ J.\  C {\bf 64}, 653 (2009)
  [arXiv:0905.3531 [hep-ph]].











\bibitem{Alekhin:2010dd}
S.~Alekhin, J.~Blumlein,P.~Jiminez-Delgado,S.~Moch and E.~Reya, arXiv:1011.6259 [hep-ph].

\bibitem{ABDV}
  R.~Bonciani, G.~Degrassi and A.~Vicini,
  JHEP {\bf 0711} (2007) 095
  [arXiv:0709.4227 [hep-ph]].\\
  U.~Aglietti, R.~Bonciani, G.~Degrassi and A.~Vicini,
  JHEP {\bf 0701} (2007) 021
  [arXiv:hep-ph/0611266].\\
  U.~Aglietti, R.~Bonciani, G.~Degrassi and A.~Vicini,
  Phys.\ Lett.\  B {\bf 595} (2004) 432
  [arXiv:hep-ph/0404071].\\
  U.~Aglietti, R.~Bonciani, G.~Degrassi and A.~Vicini,
  Phys.\ Lett.\  B {\bf 600} (2004) 57
  [arXiv:hep-ph/0407162].\\
  G.~Degrassi and F.~Maltoni,
  Nucl.\ Phys.\  B {\bf 724}, 183 (2005)
  [arXiv:hep-ph/0504137].\\
  G.~Degrassi and F.~Maltoni,
  Phys.\ Lett.\  B {\bf 600}, 255 (2004)
  [arXiv:hep-ph/0407249].

\bibitem{nnlopapers}
R.~V.~Harlander,
Phys.\ Lett.\ B {\bf 492} (2000) 74.
[arXiv:hep-ph/0007289].\\
S.~Catani, D.~de Florian and M.~Grazzini,
JHEP {\bf 0105} (2001) 025
[arXiv:hep-ph/0102227].\\
R.~V.~Harlander and W.~B.~Kilgore,
Phys.\ Rev.\ D {\bf 64} (2001) 013015
[arXiv:hep-ph/0102241]. \\
R.~V.~Harlander and W.~B.~Kilgore,
Phys.\ Rev.\ Lett.\  {\bf 88} (2002) 201801
[arXiv:hep-ph/0201206].\\
C.~Anastasiou and K.~Melnikov,
Nucl.\ Phys.\ B {\bf 646} (2002) 220
[arXiv:hep-ph/0207004].\\
V.~Ravindran, J.~Smith and W.~L.~van Neerven,
Nucl.\ Phys.\ B {\bf 665} (2003) 325
[arXiv:hep-ph/0302135].


\bibitem{LH}
  J.~R.~Andersen {\it et al.}  [SM and NLO Multileg Working Group],
  arXiv:1003.1241 [hep-ph].


\end{thebibliography}
\end{document}